\documentclass[nolongbibliography,pre,nofootinbib,twocolumn,showkeys,showpacs,preprintnumbers,amsmath,amssymb]{revtex4-2}
\usepackage{epsfig}
\usepackage{graphicx}% Include figure files
\usepackage{dcolumn}% Align table columns on decimal point
\usepackage{graphics}
\usepackage{bm}% bold math
\usepackage{subfigure}% for two column figures
\usepackage{color}
\usepackage[normalem]{ulem}
\usepackage{soul}
\usepackage{footnote}
\usepackage{url}
\usepackage{subfigure}

\usepackage{hyperref}
\usepackage{bbold}

\newtheorem{thm}{Theorem}
\newtheorem{lemma}{Lemma}
\begin{document}

\preprint{APS/123-QED}

\title{Chaotic interaction between dark matter and dark energy}  % Noooreeek!

\author{E. Aydiner}
\email{ekrem.aydiner@istanbul.edu.tr \\ https://orcid.org/0000-0002-0385-9916}

\affiliation{Theoretical and Computational Physics Research Laboratory, İstanbul University, İstanbul, Türkiye\\
Department of Physics, İstanbul University,  Fatih 34134, İstanbul, Türkiye}

%\date{Written: March 2023; Revised:  November 2023; Revised: January 2024; Revised: March 2024}

\date{Written: March 2023; Revised: March 2024}

\begin{abstract}
In this study, we consider dark matter and dark energy as grand-canonical systems which are open, non-equilibrium, coupled, and interacting systems. For the first time, we propose a new more realistic interaction scheme to explain dynamics between coupled interacting thermodynamic systems. Based on this new interaction schema, we propose new theorems to define the interactions including mutual and self-interactions. We proved the theorems based on the energy conservation law of thermodynamics. Furthermore, we obtain new coupled equations using the theorems. We numerically solved the interaction equations and obtained phase space diagrams and Lyapunov exponents. We show that the interaction between dark matter and dark energy is chaotic. We conclude that these theorems and results can be generalized to all coupled interacting particle and thermodynamic systems. Finally, we evaluated that the obtained results indicate the existence of a new law in physics. Additionally, based on the results, we gave physical definitions of basic concepts such as chaos and self-organization and pointed out the existence of hidden symmetries and variables in interacting systems.
\end{abstract}

\maketitle

%\tableofcontents

 \section{Introduction}
 \label{Intro}

It is known that our universe, which continues to expand, consists not only of matter but also of dark matter and dark energy. Dark matter, which is thought to interact with matter only through gravity, was proposed to explain the rotational anomaly of galaxies \cite{Vera1970,Rubin1980,Trimble1987}. The accelerated expansion of the universe was first discovered by Riess et al. \cite{Riess1998} by measuring luminosity distances of SNe Ia, used as standard candles. Perlmutter et al. \cite{Perlmutter1999} later confirmed the discovery with analysis of nearby and high-redshift supernovae. Estimates for the composition of the universe in terms of dark energy, dark matter and ordinary matter are given by the WMAP collaboration to be 71.4\%, 24\% and 4.6\% \cite{Bennett2013,Hinshaw2013}, and by the Planck collaboration to be 68.3\%, 26.8\% and 4.9\%  \cite{Planck2020}, respectively. 

The physical origins of dark matter and dark energy are not yet understood. This is among the important problems of the particle physics and cosmology. However, we know that these components interact at least through gravity. Therefore, the interactions between these components may play an important role in the evolution of the universe. Recently, there has been an increasing number of experimental, observational and theoretical studies aimed at understanding the physical origin of dark matter and dark energy and determining the interactions between them. For instance, there are many publications in the literature that model the interactions in the dark sector to solve singularity, cosmic coincidence, Hubble tension problems and etc
\cite{Fitch1998,Zlatev_1999,Campo2006,He2011Testing,Poitras2014,Brax2006,Bolotin2015,Zimdahl2004,Amendola2007,Wang2016,Bohmer2015,Bohmer2016,He2008,Cai2017,Yang2019,Amendola2000,Valentini2002,Amendola2003,Campo2008,Campo2009,Wei2007,He2011Testing,Poitras2014,Campo2015,Chimento2010,Sanchez2014,Verma2010,Shahalam2015,Cruz2018,Cruz2019,Saleem2020,Zhai_2023,Chatterjee_2021,Chatterjee_2022,Bhattacharya_2023,Hussain_2023,Chatterjee_2024,V_liviita_2008,Kumar_2017,Yang_2020,Kumar_2019,Murgia_2016,Di_Valentino_2021,lucca2021dark}.

In addition to the above studies,
in 2018, we proposed Chaotic Universe Theory based on interactions between the components such as matter, dark matter, and dark energy \cite{Aydiner2018}. In this study, we showed that the interaction dynamics of these components are chaotic. We suggested that the universe is evolving chaotically within the cosmic time-line. We concluded that this new cosmological model has the potential to solve many complications of the big-bang cosmology such as singularities, cosmic coincidence, emergence of galaxies, fractal formation of the universe from micro to macro scale, and the future and the end of the universe. Additionally, in another study, we showed that dark energy and dark matter interaction introduce a novel mechanism to explain late time inflation i.e., the transition from matter-dominated era to dark energy-dominated era of the universe \cite{Aydiner2022}. The first time, we obtained a hybrid scale factor containing both eras and showed that this transition appears around 9.8-10 Gyears in the cosmic time-line. Furthermore, we found that the parameters of the new hybrid scale factor are precisely consistent with the theoretical and observational results for the matter and dark energy-dominated eras. Finally, the importance of the dark matter and dark energy interactions has been thoroughly discussed in Ref.\cite{Aydiner2024ffp16}.

As we have briefly summarized above, interaction models seem to have significant potential to solve some important problems in cosmology. This is our main motivation in this study. On the other hand, in addition to this main motivation, we have several more important motivations that are the focus of this study. First of all, the nature we live in, the universe, is non-linear. This non-linearity leads to chaotic dynamics. Second, our universe has a fractal geometry from micro to macro scale. It is expected that the dynamics that lead to this geometric elegance are chaotic. This is the intuitive reason why we expect the interactions of the components in the universe to be chaotic. Third, in Ref.\,\cite{Aydiner2018}, we obtained the chaotic interactions between matter, dark matter, and dark energy inspired by Lotka-Volterra-type interactions. However, even though the Lotka-Volterra-type interaction equations of dark matter and dark energy models can cause chaotic behavior under some special conditions \cite{Aydiner2018,Perez_2014}, these equations may not fully describe the interaction between dark matter and dark energy. More importantly, we need newer approaches to the physics of interacting systems that go beyond standard approaches and reveal the deeper physics and enable understanding of nature. In this study, we will try to answer these motivations.
 
Therefore, here, we will propose a new interaction schema for dark matter and dark energy interactions, supporting the main claim of Ref.\cite{Aydiner2018} and to answer to the above motivations.
To be able to do this, firstly, we will consider dark matter and dark energy as an open and interacting thermodynamic system in a new interaction schema, unlike standard statistical thermodynamics. We have proposed two theorems as the theoretical counterpart of this interaction scheme. One of them is mutual interaction and the other is self-interaction.

Based on this new theoretical framework, we will show that interaction equations can be derived from the first law of thermodynamics. To obtain a full description of the interaction we will suggest a new complementary equation inspired by graph theory. Finally, based on these theorems, we will carry out new interaction equations that describe the interaction between dark matter and dark energy. We will show that the interaction between dark matter and dark energy 
is chaotic. On the other hand, we aim to propose a new law of physics, inspired by the possibility that the obtained results may be universal for all thermodynamic systems. Additionally, we will introduce and explore several new conceptual definitions.
We believe that our results might give a critical contribution not only to cosmology but also to statistical thermodynamics and the understanding of nature and living systems.

This work is organized as follows: In Section \ref{Conventional}, we briefly summarize the conventional method for modeling the interaction between dark matter and dark energy, in Section \ref{New_Theory}, we introduce a new and novel interaction schema to model dark matter and dark energy interactions. We present new theorems and give proofs based on the energy conservation law of thermodynamics. We obtain new interaction equations between coupled non-equilibrium systems. In Section \ref{Numerical}, the numerical solution of the new model is given. We present some remarkable conclusions in Section \ref{Discussion}. We sum up the important results in Section \ref{Summary}. Finally, in the last Section \ref{Further}, we point out possible future studies.

\section{The Conventional Approach to the Interaction} 
\label{Conventional}

Assuming the coexistence of dark matter and dark energy as components in the Universe, the continuity equation for both these components is given by
\begin{equation} \label{non-int-continutiy}
	\dot{\rho} +  3 H \left(\rho + p \right)   = 0 .
\end{equation} 
If we assume that they interact in the same volume, the continuity equation (\ref{non-int-continutiy}) can be written for the dark matter and dark energy separately as
\begin{subequations} \label{couple-dm-de}
	\begin{equation}
		\frac{d \rho_{dm}}{dt}  +  3 H \left( \rho_{dm} + p_{dm} \right)   = X
	\end{equation}
	\begin{equation}
		\frac{d \rho_{de}}{dt}  +  3 H \left( \rho_{de} + p_{de} \right)   = -X
	\end{equation}
\end{subequations}
where $X$ is the interaction coupling term. 

So far many different interaction models have been proposed in literature. For example, in our previous study \cite{Aydiner2018}, we consider $X=\gamma \rho_{dm} \rho_{de}$ where $\gamma$ is the interaction constant. By setting $\alpha_{1}= 3H(1+w_{dm})$,  $\alpha_{2}=-3H(1+w_{de})$, $x_{1}=\frac{\gamma}{\alpha_{2}} \rho_{dm}$ and $x_{2}=\frac{\gamma}{\alpha_{1}} \rho_{de}$, 
Eq.(\ref{couple-dm-de}) can be written in terms of Lotka-Volterra equation \cite{Aydiner2018}:
\begin{subequations} \label{LV}
	\begin{equation} 
		\frac{d x_{1}}{dt}  = \alpha_{1} x_{1} \left(  1 - x_{2}   \right)
	\end{equation}
	\begin{equation}
			\frac{d x_{2}}{dt}  = \alpha_{2} x_{2} \left(  x_{1} - 1   \right)
	\end{equation}
\end{subequations}
where $\alpha_{1,2}$ are the interaction rate parameters.

The phase space solution is periodic of the Lotka-Volterra equation in Eq.(\ref{LV}). Based on this idea, in the previous study \cite{Aydiner2018}, Eq.(\ref{LV}) was extended to a more general interaction by adding quadratic terms representing the self-interaction,
	\begin{equation} 
	\frac{d x_{i}}{dt}  =  x_{i}  \sum_{j}^{N}  \alpha_{ij} \left(  1 - x_{j},   \right) \qquad i=1,...,N
\end{equation}
which is the same expression as in Refs. \cite{Arneodo1980occurence,Vano2006Chaos}. It was shown that the solutions obtained from the coupled density equations already show that the dynamics of the system is chaotic. In his approach, different components or kinds are consuming each other. However, this approach may violate the conservation laws. Therefore, we say that this interaction method is incomplete from the point of view of statistical thermodynamics. We will present a new method that covers the full thermodynamic behavior of the interacting system.

\section{New Interaction Theory}
\label{New_Theory}

\begin{figure} [h!]
\centering
	\includegraphics[scale=0.8]{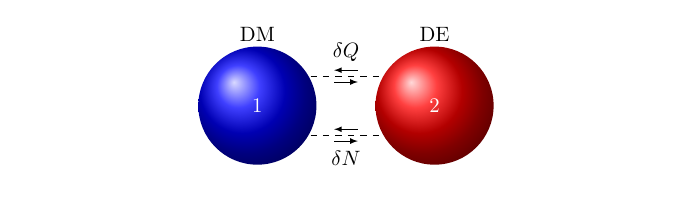}
	\caption{The schematic picture for the thermodynamic ensemble of the dark matter (DM) and dark energy (DE).  In the standard thermodynamic schema, open thermodynamic ensembles interact via the $\delta Q$ and $\delta N$ exchange. Here we set $\delta V=0$.}
	\label{xfig1}
\end{figure}

We consider dark matter and dark energy as grand canonical thermodynamic systems. According to the conventional approach, they interact in the Universe by exchanging energy and particles as seen in Fig.\ref{xfig1}. 
However, we show that the conventional approach is not enough to define the interactions between two open, coupled non-equilibrium thermodynamic systems. Therefore, we introduce new interaction schemes and theorems to define the correct interaction equations. Below, for simplicity, we label dark matter and dark energy as systems $1$ and $2$, respectively.

\subsection{Mutual Interactions}

\begin{lemma}
The continuity equation for a non-equilibrium system $1$ which interacts with the system $2$ should be presented in the form as
	\begin{equation}  \label{n-continuty}
		\frac{d \rho_{1}}{dt}  +  3 H \left( \rho_{1} + p_{1} \right)   = \frac{d X}{dt}
	\end{equation}
where $X$ is the source term of system $2$ based on the approach of Prigogine \cite{Prigogine1988}.
\end{lemma}
Change of the source $X$ can be defined as $X= \Gamma_{1} x_{2}$ where $x_{2}$ is the energy flux from system 1 to the 2, $\Gamma_1$ denotes the \textit{energy transfer rates}, i.e., the system $1$ which determines the energy amount which is transferred the from system $2$ to the system $1$. In the present study, based on this Lemma, we present a new interaction schema and new theorems. 

\begin{thm} \label{theorem-1}
The interactions between two coupled non-equilibrium systems can be obtained from the energy conservation law of statistical thermodynamics and is given in a simple form as
\begin{subequations}  \label{Aydiner-eq-1}
	\begin{eqnarray} 
		\frac{d x_{1}}{dt}  = \Gamma_{1} x_{2} -   x_{1}  \label{Aydiner-1}
	\end{eqnarray}
	\begin{eqnarray}
		\frac{d x_{2}}{dt}  =  \Gamma_{2} x_{1}  - x_{2}  \label{Aydiner-2}
	\end{eqnarray}
\end{subequations}
where $x_{1,2}$ denotes the energy flux between systems and $\Gamma_{1,2}$ the energy transfer which are time-dependent degrees of freedom of the system. 
\end{thm}

\textit{\textbf{Proof}:}  The two open and coupled interacting 
systems $1$ and $2$ are given in Fig.\ref{xfig1}. In the standard statistical thermodynamics, the open and interacting two systems are characterized by the energy and particle exchange as in Fig.\ref{xfig1}. The energy conservation law of thermodynamics for open thermodynamic systems is given by
	\begin{eqnarray}
	dU = \delta Q  - p \delta V  + \mu \delta N
\end{eqnarray}
where $dU$ corresponds to the energy exchange in the system, $\delta Q$ is the change of heat, $p$ is the pressure, $\delta V$ denotes volume exchange, $\mu$ is the chemical potential and $\delta N$ is the particle exchange in the same system. 

For the constant volume $\delta V=0$, the energy conservation laws of the thermodynamics for two non-interacting systems are given separately as 
\begin{subequations}
	\begin{eqnarray}
		dU_{1} =   \delta Q_{1}  + \mu_{1} \delta N_{1} 
	\end{eqnarray}
	\begin{eqnarray}
		dU_{2} =  \delta Q_{2}  + \mu_{2} \delta N_{2} 
	\end{eqnarray}
\end{subequations}
These equations can be represented as a time-dependent form
\begin{subequations}  \label{internal-energy} 
	\begin{eqnarray}
		\frac{dU_{1}}{dt} =   \frac{dQ_{1}}{dt}  + \mu_{1}  \frac{dN_{1}}{dt} 
	\end{eqnarray}
	\begin{eqnarray}
		\frac{dU_{2}}{dt} =   \frac{dQ_{2}}{dt}  +  \mu_{2}  \frac{dN_{2}}{dt} 
	\end{eqnarray}
\end{subequations}
It should be noted that these equations do not represent the interaction between systems. Instead, they denote only independent energy change in two systems in contact with each other. It is a conventional mistake to say that the interactions between two thermodynamic systems are given by Eq.(\ref{internal-energy}).

\begin{figure} [h!]
	\centering
	\includegraphics[scale=0.8]{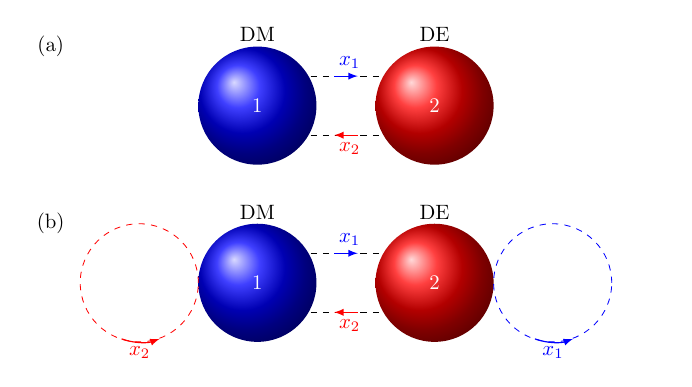}
	\caption{A new interaction schema. In (a) $x_{1}$ and $x_{2}$ denote the mutual interactions between two systems, in (b) loops given dashed circles represent the self-interaction of each ensemble. }
	\label{xfig2}
\end{figure}

To write the interaction between two systems, it is necessary to accept that the two systems couple. The interaction naturally connects the two systems and the flow of energy occurs between the two systems. The energy flow between the two systems must be bidirectional, as shown in Fig.\ref{xfig2}(a), due to the nature of the interaction. This exchange represents \textit{mutual interaction.} The energy fluxes $x_{1}$ and $x_{2}$ in Fig.\ref{xfig2}.(a) can be determined by Eq.(\ref{internal-energy}). As we will show below these fluxes represent the changes in $\delta Q$ and $\delta N$ between the two systems in terms of energy. However, it is not enough to assume that the energy flow is bidirectional as in Fig.\ref{xfig2}(a).  Energy flow cannot be written independently of the source. Therefore, as \textbf{Lemma} 1 indicates, the flow must be written depending on the source. When the energy flow is written as dependent on the source, it also makes sense in the change that occurs on the source. The change in the source is represented as \textit{self-interaction}. The self-interaction loops that appear on the sources are indicated with a dashed circle in Fig.\ref{xfig2}(b). This surprising connection shows that the existence of mutual interaction depends on self-interaction.

The new interaction schema in Fig.\ref{xfig2}(b) is more realistic and introduces a new perspective to the model for coupling systems. The main problem in this new interaction scheme is how to incorporate this self-interaction contribution into the system. Here, we will establish this connection when establishing the mutual interaction equations. However, we will also discuss the self-interaction contribution separately. We only state that these self-interaction loops can be considered as \textit{hidden actions} including \textit{hidden vectorial variables}. 

By using \textbf{Lemma} 1, the internal energy change in Eq.(\ref{internal-energy}) can be written as
\begin{eqnarray} \label{exchange-U1-U2}
	\frac{dU_{1}}{dt} =  M_{1} \Gamma_{1} x_{2}, \qquad \frac{dU_{2}}{dt} =  M_{2} \Gamma_{2} x_{1}
\end{eqnarray} 
where $M_{1,2}$ is the mutual interaction coupling constant and $\Gamma_{1,2}$ are the \textit{energy storage speed} of each of the systems for the cyclic dynamic in Fig.\ref{xfig2}(b). $\Gamma_{1,2}$ are \textit{the non-holonomic variables}. 

Eq.(\ref{exchange-U1-U2}) clearly says that the internal energy change in system 1 is determined by system 2, and conversely, the internal energy change in system 2 is determined by system 1.

On the other hand, the first term on the right-hand side of Eq.(\ref{internal-energy}) denotes energy change in each of the systems due to coupling. Therefore the heat $Q_{1,2}$ can be defined as the heat flux and it can be written as $Q_{1,2} = \phi_{1,2} =  L_{1,2} x_{1,2}$. Therefore, the heat terms are given by
\begin{eqnarray}  \label{energy-store}
	\frac{dQ_{1}}{dt} =  L_{1} \frac{d x_{1}}{dt}, \qquad 	\frac{dQ_{2}}{dt} =  L_{2} \frac{d x_{2}}{dt}.
\end{eqnarray} 
where $L_{1,2}$ are constants and indicate \textit{the energy store capacity of the system}. Then, the second term on the right-hand side of Eq.(\ref{internal-energy}) can be regarded as the changing of the energy amount with time. Therefore, the second term can be written as
\begin{eqnarray}  \label{particle-store}
	\mu_{1}  \frac{dN_{1}}{dt} =  \mu_{1}   x_{1}, \qquad 	\mu_{2} \frac{dN_{2}}{dt} =  \mu_{2} x_{2} .
\end{eqnarray} 

Finally, to simplify equations we set $ L_{1}= L_{2}=1$, $ \mu_{1}= \mu_{2}=1$ and $M_{1}=M_{2}=1$. If we insert Eqs.(\ref{exchange-U1-U2}), (\ref{energy-store}) and (\ref{particle-store}) into Eq.(\ref{internal-energy}), we can obtain interaction equations in Eq.(\ref{Aydiner-eq-1}) as
%\begin{subequations} % \nonumber % \label{Aydiner-equations-1} 
	\begin{equation}  
		\frac{d x_{1}}{dt}  =  \Gamma_{1} x_{2} -   x_{1}   \tag{\ref{Aydiner-1}} 
	\end{equation}
	\begin{equation}
		\frac{d x_{2}}{dt}  =  \Gamma_{2} x_{1}  - x_{2}  \tag{\ref{Aydiner-2}}
	\end{equation}
%\end{subequations}
which prove \textbf{Theorem} \ref{theorem-1}. 

Here, we introduce the mutual interaction schema and show that the mutual interaction equation between dark matter and dark energy can be obtained from the energy conservation laws of thermodynamics.  However, this description in Eq.(\ref{Aydiner-eq-1})  is also not enough to define the interaction dynamics of two interacting and cyclic non-equilibrium systems. To obtain a completed interaction schema,  non-holonomic vectorial variables $\Gamma_{1}$ and $\Gamma_{2}$ and contribution of the self-interaction loops should be defined.

\textbf{Theorem} \ref{theorem-1} imply that given system in Fig.\ref{xfig2}(b) to be cyclic.  The presence of cyclic interaction and the presence of vectorial non-holonomic variables $\Gamma_{1,2}$ indicate that there must be other driving vectorial forces in the system. Without these vectorial forces, it may not be possible to drive the system. At this point, we can ask the main questions:  \textit{How to define the non-holonomic time-dependent degrees of freedom of the system?} and \textit{What are these hidden vectorial fields?} To reply to these questions and to complete the interaction description we have to deal with the role of the self-interacting loops.

\subsection{Self-Interactions}

\begin{thm}\label{theorem-2}
The non-holonomic time-dependent energy transfer rate parameters $\Gamma_{1,2}$ can be obtained from hidden
vectorial fields and are given in terms of  $x_{1}$ and $x_{2}$ currents as 
\begin{subequations}  \label{Aydiner-eq-2}
	\begin{equation} 
	\frac{d \Gamma_{1}}{dt} = 1 - x_{1} x_{2}     \label{ata-3}
\end{equation} 
\begin{equation}  
	\frac{d \Gamma_{2}}{dt} =  1 - x_{1} x_{2}     \label{ata-4}
\end{equation} 
\end{subequations}
\end{thm}

\begin{figure} [h!]
	\centering
	\includegraphics[scale=0.8]{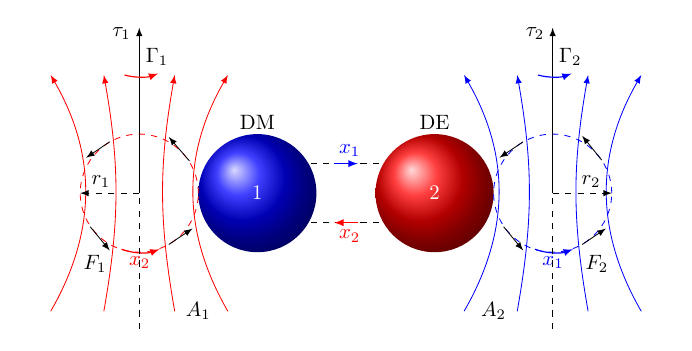}
	\caption{The vectorial attractor fields and torque behavior on the self-interacting loops. $\Gamma_{1,2}$ denotes non-holonomic variables in Eq.(\ref{Aydiner-eq-1}). $\vec{A}_{1,2}$ are the attractor vectorial fields caused by driven force $\vec{F}_{1,2}$ and $\vec{\tau}_{1,2}$ correspond to torque on the loops.}
	\label{xfig3}
\end{figure}

\textit{\textbf{Proof}:} Before giving a mathematical proof, it will be useful to give the following definitions and descriptions to explain the dynamics that arise in the self-interaction scheme:

\begin{itemize}
	\item Self-interaction loops given in Fig.\ref{xfig2}(b) represent the dynamics within systems 1 and 2.
	
	\item As seen from Fig.\ref{xfig3}, the energy $x_{1}$ flows from system 1 to system 2 and acts on system 2. Similarly, $x_{2}$ flows from the system 2 to the system 1 and acts on the system 1. While  $x_{1}$ causes a cyclic dynamics within the system 2, and $x_{2}$ also causes a cyclic dynamics on the system 1.
	
	\item Energy flow $x_{2}$ in system 1 changes $\Gamma_{1}$ in the system 1, however, $\Gamma_{1}$ also changes the energy flow $x_{2}$. Therefore, it can be suggested that there is a non-linear relationship between these two variables. A similar situation also applies to the dynamics in system 2. 
	
	\item These cyclic dynamics of $x_{1,2}$  on the systems produce the non-holonomic vectorial variables which are represented by $\Gamma_{1,2}$ as seen in Fig.\ref{xfig3}. 
	
	\item The rotational flow of $x_{1,2}$ on the systems will cause inertial momentum in both systems. 
		
	\item Energy flow $x_{1}$  and $x_{2}$ on the systems leads to the vectorial forces $\vec{F}_{1,2}$. 
	
	\item The cyclic dynamics on the loops lead to torques $\vec{\tau}_{1,2}$ we call them as \textit{vectorial attractor torques}.
	
	\item The energy flows $x_{1,2}$ can be written in terms of the vectorial force $\vec{A}_{1,2}$ as $\vec{\nabla}\times \vec{A}_{1,2} \sim  \vec{x}_{1,2} $. We call these forces as  \textit{vectorial attractor fields}.
	
	\item Except the energy flows $x_{1,2}$, all vector quantities occurring on the self-interaction loops and driving the systems can be regarded as virtual or pseudo quantities, we called these quantities \textit{hidden}.
		
\end{itemize}

After these definitions and descriptions, we can discuss the relation between these vectorial quantities to describe the dynamics of the self-interaction loops. It is possible to interconnect all virtual vector quantities of the self-interaction loops within the torque term.

\textit{The vectorial attractor torque} can be written in terms of \textit{energy transfer velocity} as
\begin{subequations}  \label{tork-angular-velocity}
	\begin{eqnarray}  \label{tork-vel-1}
		\vec{\tau}_{1} = c_{1} \frac{d \Gamma_{1}}{dt}  \hat{z} =   \vec{\tau}_{1b} + 	\vec{\tau}_{1f} 
	\end{eqnarray} 
	\begin{eqnarray}  \label{tork-vel-2}
	   \vec{\tau}_{2} = c_{2} \frac{d \Gamma_{2}}{dt}  \hat{z} =  \vec{\tau}_{2b} + 	\vec{\tau}_{2f} 
	\end{eqnarray} 
\end{subequations}
where $c_{1,2}$ are inertia momenta of the system 1 and 2, $ \vec{\tau}_{1b,2b} $ denote torques belong to the bulk and $\vec{\tau}_{1f,2f}$ are  torque cause from flux $x_{1,2}$. 

The bulk torques can be assumed constant and do not contribute to the formation of the flux $x_{1,2}$, therefore, can be ignored or assigned constant values. The main torques $\vec{\tau}_{1f} $ and  $\vec{\tau}_{2f}$ an the loop 1 and 2 cause form $d\vec{F}_{1}$ and $d\vec{F}_{2}$, respectively. These torques can be written in terms of vectorial fields $\vec{A}_{1}$ and $\vec{A}_{2}$ as 
\begin{subequations} \label{tork}
\begin{eqnarray}  \label{tork1}
	\vec{\tau}_{1f} = \int    d\vec{F}_{1} \times \vec{r}_{1}  = x_{1} \int \left( d \vec{r}_{1} \times \vec{A}_{1}  \right) \times  \vec{r}_{1} 
\end{eqnarray} 
\begin{eqnarray}  \label{tork2}
	\vec{\tau}_{2f} = \int  d\vec{F}_{2} \times \vec{r}_{2}  = x_{2} \int \left( d \vec{r}_{2} \times \vec{A}_{2}  \right) \times  \vec{r}_{2} 
\end{eqnarray} 
\end{subequations}
where loop radius $r_{1,2}$ characterize the velocity magnitude of the densities and $A_{1,2}$ are the vectorial fields on the loops.  The integrals in Eq.(\ref{tork}) can be written as 
\begin{subequations} \label{tork-int}
	\begin{eqnarray}  \label{tork1-1}
	\int \left( d \vec{r}_{1} \times \vec{A}_{1}  \right) \times  \vec{r}_{1} = -  M_{1}  x_{2}  \hat{z}
	\end{eqnarray} 
	\begin{eqnarray}  \label{tork2-2}
	 \int \left( d \vec{r}_{2} \times \vec{A}_{2}  \right) \times  \vec{r}_{2} = -  M_{2} x_{1}   \hat{z}
	\end{eqnarray} 
\end{subequations}
where $M_{1,2}$ are the positive constants. As a result, combining Eqs.(\ref{tork-angular-velocity}), (\ref{tork}) and (\ref{tork-int}), we obtain
\begin{subequations} \label{tork-int-const}
	\begin{eqnarray}  \label{tork1-c}
	  c_{1} \frac{d \Gamma_{1}}{dt} =	 \tau_{1b}  -  M_{1} x_{1}  x_{2}
	\end{eqnarray} 
	\begin{eqnarray}  \label{tork2-c}
	  c_{2} \frac{d \Gamma_{2}}{dt} =  \tau_{2b} -  M_{2} x_{1} x_{2}
	\end{eqnarray} 
\end{subequations}
Now, if we set $ c_{1} = c_{2} =1$ and $\tau_{1b}=\tau_{2b}=1$, we obtain equations for non-holonomic variables $\Gamma_{1,2}$
as
%\begin{subequations}  % \nonumber
\begin{equation}  \nonumber
    \frac{d \Gamma_{1}}{dt} =  1 - x_{1} x_{2}   \tag{14a} %  \tag{\ref{ata-3} revisited}
\end{equation} 
\begin{equation}   \nonumber
    \frac{d \Gamma_{2}}{dt} =  1 - x_{1} x_{2}    \tag{14b}   % \tag{\ref{ata-4} revisited} 
\end{equation} 
%\end{subequations}
which verify \textbf{Theorem} \ref{theorem-2}. Here we show that the density $x_{1}$ and $x_{2}$ behave like current for both systems. 
The currents $x_1$ and $x_2$ in the systems are described by vector fields that govern the interaction scheme in the cyclic dynamics.

\subsection{ Full Interactions}

As a result, based on \textbf{Theorems}~\ref{theorem-1} and ~\ref{theorem-2}, full interaction between dark matter and dark energy can be given by
	\begin{eqnarray}  \label{Aydiner}
		\frac{d x_{1}}{dt}  = x_{2} x_{3} -  x_{1}  \nonumber \\
		\frac{d x_{2}}{dt}  =  (x_{3}- q)  x_{1}  - x_{2} \\
		\frac{dx_{3}}{dt} = 1 - x_{1} x_{2} \nonumber
	\end{eqnarray} 
where $q$ is the control parameter which can be given as $q=\Gamma_{1}-\Gamma_{2}$  \cite{[{Similar equations were derived in a different context in}]  rikitake1958oscillations}. These are non-linear coupled equations that govern the dynamics of the dark energy and dark matter interactions in Fig.\ref{xfig1}. 

\section{Numerical Results}
\label{Numerical}

\begin{figure} [h!]
	\centering
	\includegraphics[height=6cm,width=8cm]{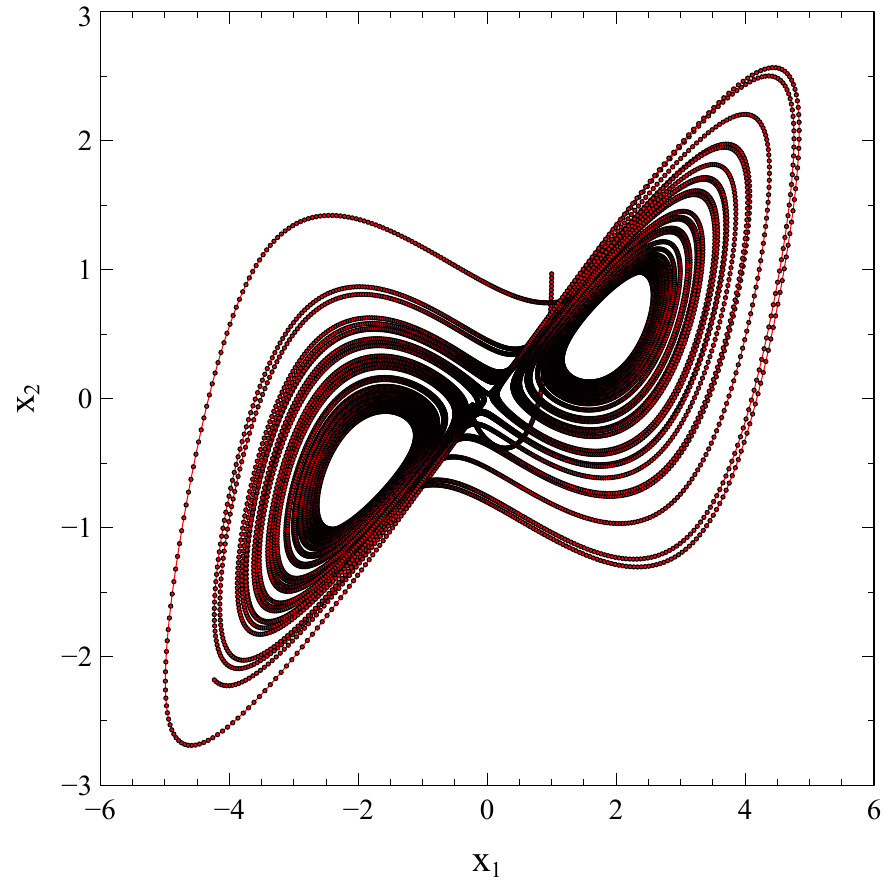}
	\caption{Chaotic attractor and trajectory in $x_{1}-x_{2}$ plane for the control parameter $q=3.46$.}
	\label{phase-1}
\end{figure}

\begin{figure} [h!]
 		\centering
 		\includegraphics[height=6cm,width=8cm]{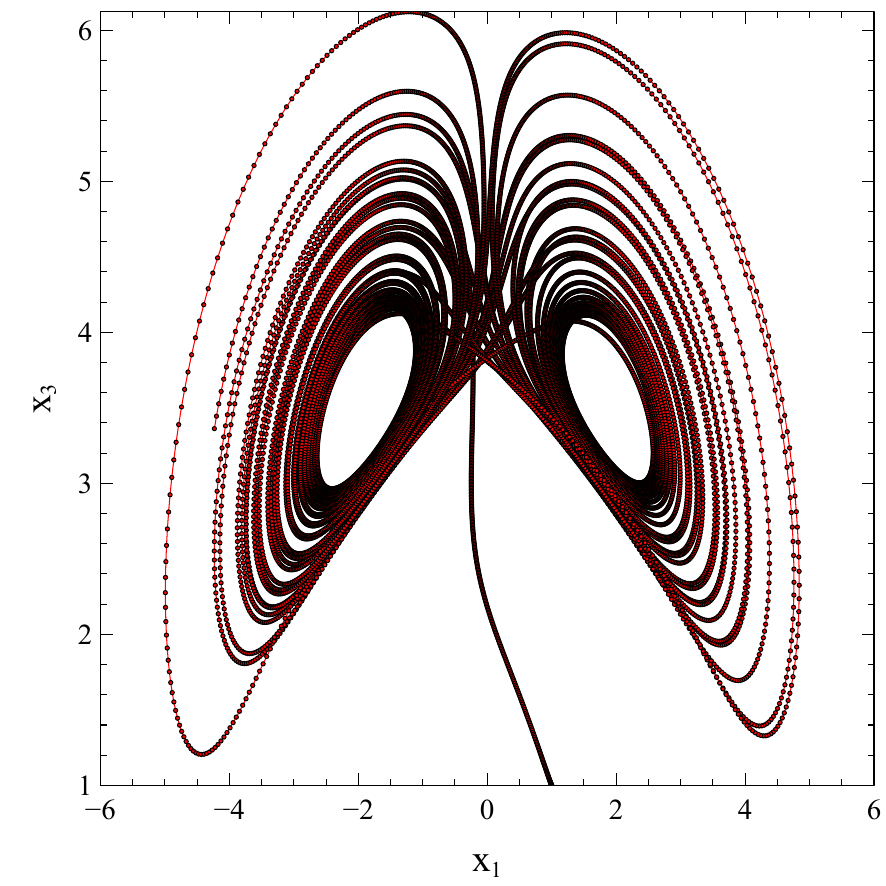}
 		\caption{Chaotic attractor and trajectory in $x_{1}-x_{3}$ plane for the control parameter $q=3.46$. }
 		\label{phase-2}
 	\end{figure}

\begin{figure} [h!]
	\centering
	\includegraphics[height=6cm,width=8cm]{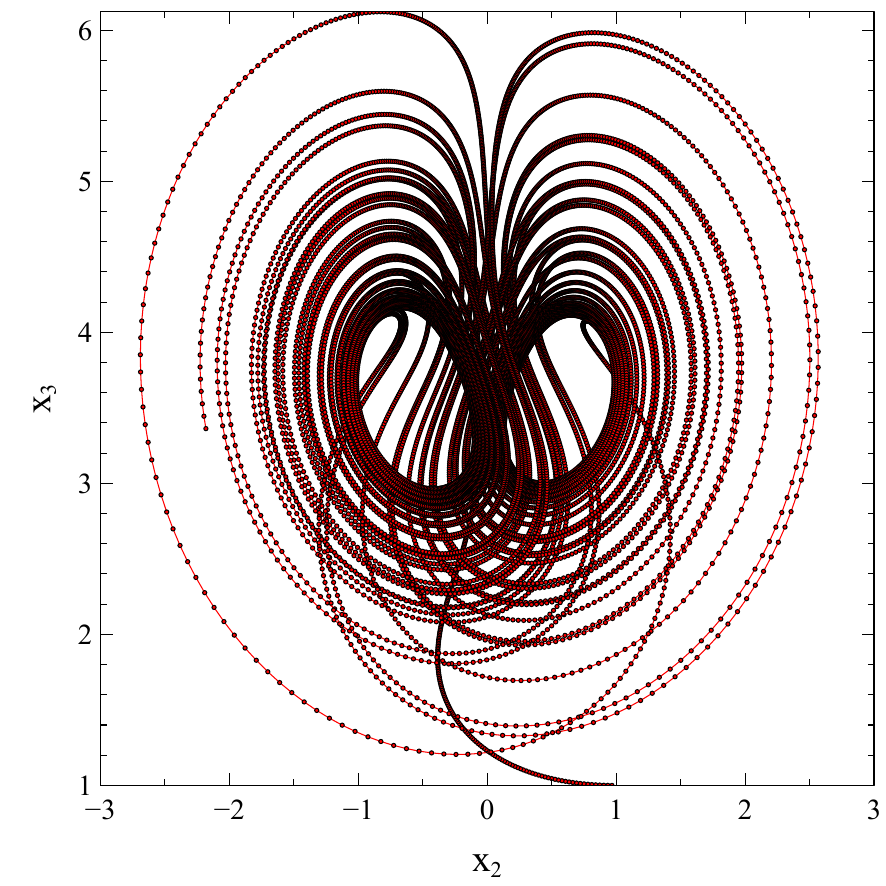}
	\caption{Chaotic attractor and trajectory in $x_{2}-x_{3}$ plane for the control parameter $q=3.46$.}
	\label{phase-3}
\end{figure}

To analyze the dynamical behavior of Eq.(\ref{Aydiner}), we solved these equations numerically by writing FORTRAN 90 code based on the Runga-Kutta method and the linearized algorithm \cite{Wolf1985}. Phase trajectories are given in Figs.\ref{phase-1}-\ref{phase-3}. 

In Fig.\ref{phase-1}, the trajectory is given in the $x_1-x_2$ phase plane for the control parameter $q=3.46$. In this figure, there is an attractor around $x_{2}=0$, and the trajectory proceeds to the right without cutting itself.

 In Fig.\ref{phase-2}, the attractor appears more clearly in $x_{1}-x_{3}$ plane. In this figure, it can be seen that the center of the attractor is located around $x_{3}=0$. As seen from Fig.\ref{phase-2} that trajectory proceeds to the right in direction $x_{1}$ without cutting itself.
 
In Fig.\ref{phase-3}, the trajectory is given in the $x_{2}-x_{3}$ phase plane for the control parameter $q=3.46$. In this figure, there is an attractor around $x_{2}=0-0.5$ and at $x_{3}=1$. 

\begin{figure} [h!]
	\centering
	\includegraphics[height=8cm,width=8cm]{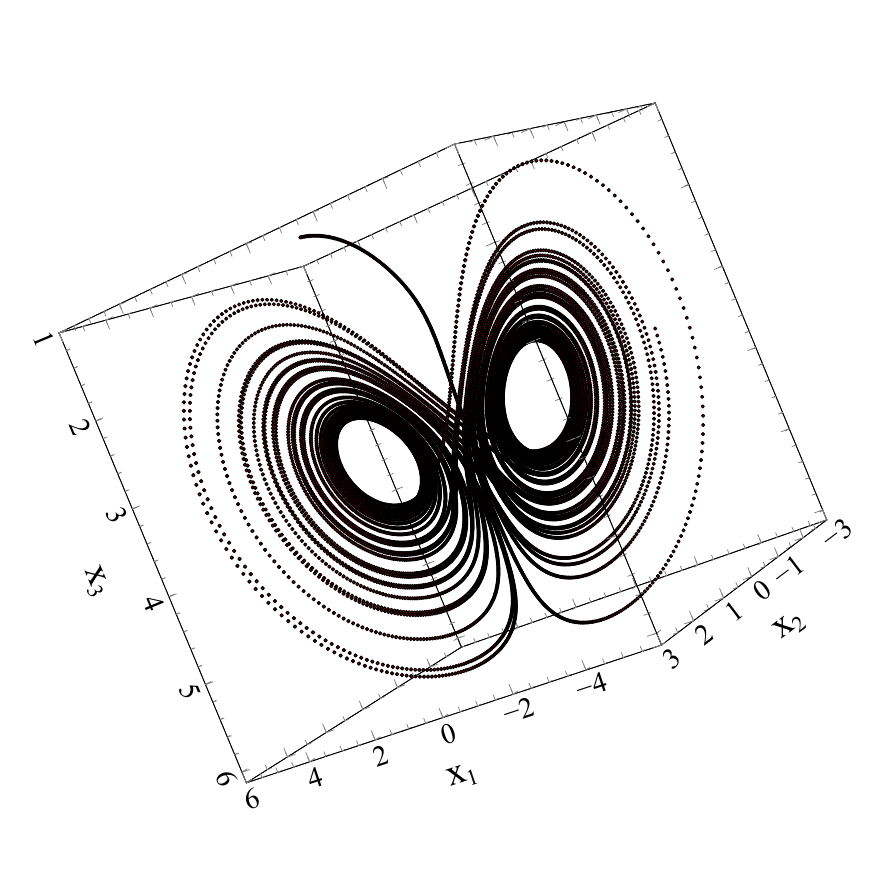}
	\caption{Chaotic attractors in $x_{1}-x_{2}-x_{3}$ phase plane for the control parameter $q=3.46$.}
	\label{3dphase-3}
\end{figure}
This situation can be observed more clearly in Fig.,\ref{3dphase-3}. In this phase space, $x_{1}$, $x_{2}$, and $x_{3}$ are plotted against each other for the same initial conditions and $q$ value. From Fig.,\ref{3dphase-3}, it is evident that the trajectories within the attractors do not intersect. If the trajectories were to intersect, the solutions would no longer be expected to exhibit chaotic behavior. The crucial point here is the demonstration of the existence of chaotic attractors. It is possible to reproduce such plots for various $q$ values. Whether these attractors are indeed chaotic can be further clarified using the Lyapunov exponent.

\begin{figure} [h!]
	\centering
	\includegraphics[height=6cm,width=8cm]{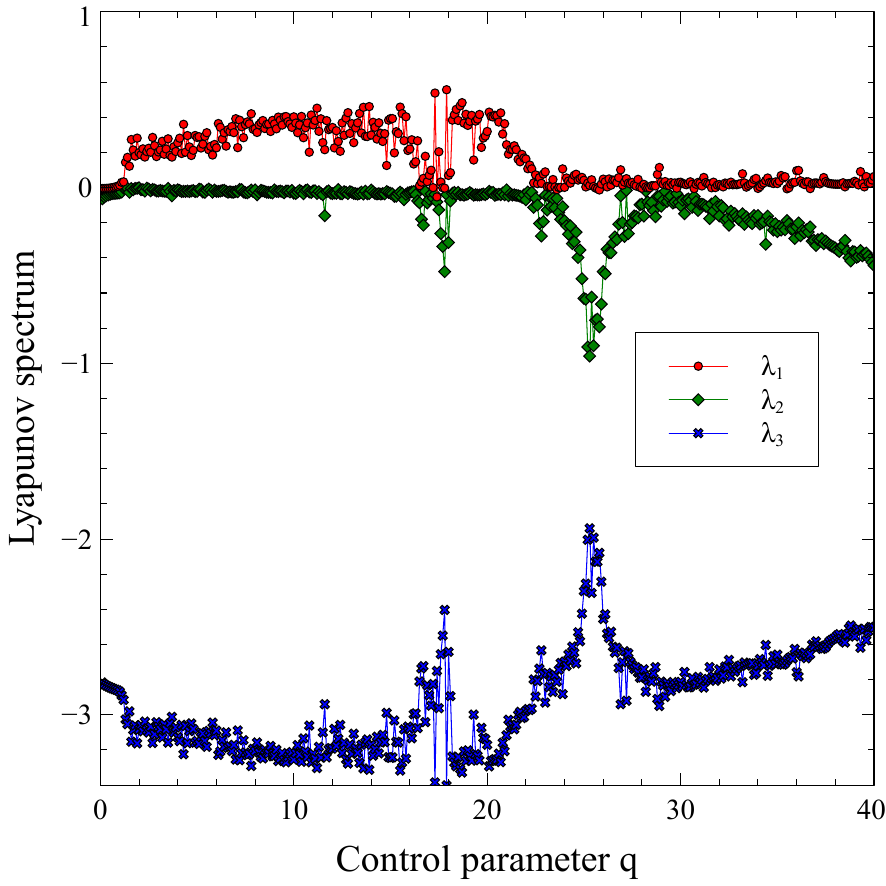}
	\caption{Lyapunov spectrum versus control parameter $q$ for the DM-DE interaction. }
	\label{q-1-lyap}
\end{figure}

Additionally, we compute the Lyapunov exponent by using the Wolf algorithm and give the numerical results in Fig.\ref{q-1-lyap}. 
As seen from Fig.\ref{q-1-lyap}, there are three Lyapunov exponents $\lambda_{i}$, $i=1,2,3$ for the control parameter $q$. Positive Lyapunov exponents indicate the chaotic behavior of the system. One can see that two exponents (red and green curves) 
have positive values depending on the control parameter $q$.
Lyapunov spectrum in Fig.\ref{q-1-lyap} shows that the system in Eq.(\ref{Aydiner}) behaves chaotic. Based on these numerical results we conclude that the interaction between dark matter and dark energy is chaotic.

It can be observed from Fig.\ref{q-1-lyap} that the Lyapunov exponent has a positive value for $q<25$, while it fluctuates around zero for $q>25$. A more detailed analysis may be required to examine the specifics; however, this is not essential for the current discussion. It is sufficient to note that a positive Lyapunov exponent indicates chaotic behavior in the system. Nevertheless, the system may not exhibit chaos for all values of $q$. The parameter $q$ represents the difference between $\Gamma_{1}$ and $\Gamma_{2}$. As this difference increases beyond $q>25$, the system may transition out of chaotic behavior. Specifically, scenarios where $\Gamma_{1}$ $\gg$ $\Gamma_{2}$ or $\Gamma_{1}$ $\ll$ $\Gamma_{2}$ could lead the nonlinear system to produce linear solutions. Consequently, the interaction between dark matter and dark energy would no longer exhibit chaotic dynamics.
To analyze this phenomenon from a cosmological perspective, it is crucial to determine the appropriate range of $q$. Achieving this requires precise fitting of parameters such as the dark matter and dark energy densities, the equation of state parameter $\omega$, cosmological quantities like the scale factor $a(t)$, and the known data from the $\Lambda$CDM model. This topic will be revisited in Section \ref{Further}.

In this study, it was shown for the first time that the interactions between dark matter and dark energy should be universally chaotic. These results are not only limited to dark matter and dark energy interactions but also have a universality to be generalized for all interacting particle and thermodynamic systems.

\section{Discussion and Some Concluding Remarks} \label{Discussion}

In this study, firstly, we briefly summarized the conventional approach to the dark matter and dark energy interaction. We argued that equations of the conventional approach may not fully reflect the interaction between dark matter and dark energy, since the interaction terms are chosen arbitrarily.

We propose a new interaction scheme that represents the interaction of dark matter and dark energy. We have demonstrated that interactions should be defined in two ways. The first is the mutual interaction between the two systems and the second is the self-interaction that occurs on each system. These two interaction pictures are beyond what has been known so far and offer a new approach to the interaction mechanisms of interacting systems. Based on the new interaction schema, we introduced several new concepts such as \textit{energy transfer velocity}, \textit{vectorial attractor fields}, \textit{vectorial attractor torques} which are new concepts for the non-equilibrium statistical mechanics and thermodynamics. Furthermore, we introduced new theorems to define the mutual and self-interactions.

In \textbf{Theorem}~\ref{theorem-1}, we introduce a new coupling equation to define the mutual interactions between dark matter and dark energy. We give a proof for this theorem based on the energy conservation law of thermodynamics. We also state that equations in \textbf{Theorem}~\ref{theorem-1} include non-holonomic variables. We know that two coupled systems out of equilibrium will exchange energy and particles until they reach fundamental and chemical equilibrium. However, the system we give in Fig.\ref{xfig2} is not just an out-of-equilibrium coupled system. Since we assume that these systems are completely transformed into each other, the dynamics of the system are not spontaneous. The system is startled by the forces controlling the interaction dynamics.
Therefore, we need to define some vectorial forces to provide the transformation of dark matter to transform into dark energy or vice versa. However, these vectorial forces drive $x_{1}$ to $x_{2}$ or $x_{2}$ to $x_{1}$ in this dynamics.

In \textbf{Theorem} \ref{theorem-2}, we introduce a new complementary coupling equation to define the dynamics of the non-holonomic variables. We also give a proof for \textbf{Theorem}~\ref{theorem-1} based on a self-interacting loop in Fig.\ref{xfig3} motivated by the node connections approach of the graph theory.

By using these theorems, we obtain a non-linear coupling equation that fully describes dark matter and dark energy interactions. 
We numerically solve these equations and we obtain phase space trajectories in various phase planes. We compute Lyapunov exponents for the control parameter $q$ and show that there are non-negative values of the exponents which indicates chaos. It is the first time, based on proof-able theorems and new non-linear coupling equations, that we show that the interaction between dark matter and dark energy is fully chaotic. Here it should be noted that, until now, in physics, changes in thermodynamic systems in contact with each other were interpreted as interactions. However, these changes do not represent mutual interaction. It is necessary to write the interactions depending on the source. When interactions are written depending on the source, mutual interaction emerges. The basic philosophy of this study is based on this idea. In this study, the concept of mutual interaction is proposed for the first time for thermodynamic systems that come into contact with each other. As a result, the mutual interaction causes self-interaction in systems. In other words, self-interactions are the result of mutual interactions. If there are no mutual interactions, there will be no self-interaction. 

It must be admitted that it is easy to visualize the mutual interaction. For example, if there is a heat transfer $\delta Q$
from the other system to the first one, it can be assumed that heat $-\delta Q$ is transferred from the other system to the other.
However, self-interaction may not be easy to understand or explain, because no one would expect such an interaction mechanism in the classical thermodynamic system. It can be thought that since such an effect has not been expected yet, it has not been theorized.  However, this counter-intuitive assumption may be understandable upon careful consideration. It should be kept in mind that if energy from one system goes to another system and causes dynamics within that system, there must be a mechanism that sets this dynamic. In this study, we propose that this mechanism can be established by the so-called forces that arise in the system. In conclusion, the interaction model we present here is quite interesting and new. Thanks to this model, it may be possible to understand the physical nature of interacting systems. Although the model seems contrary to common sense, for now, it should not be overlooked that it has the potential to add new definitions and concepts to physics.
 
Furthermore, one can see that these significant results can be generalized to all interacting systems such as matter and dark matter interaction visa versa. Our findings shed light on a more accurate understanding of the dynamics of the components in the universe. Furthermore, these results strongly prove that the universe has chaotic dynamics. In the previous study \cite{Aydiner2018}, we proposed that interactions between matter, dark matter, and dark energy would be chaotic. At the same time, we stated that the universe evolves through chaotic interactions and that the universe has cosmic evolution with cyclic chaotic processes. In this study, we have strongly proved these hypotheses in a new theoretical framework.

This theoretical evidence not only explains the interaction of dark matter and dark energy but also reveals an important and striking perspective in terms of understanding how nature works. The interaction equations we have obtained have provided a completely new understanding of nature by illuminating the hitherto unnoticed behavior of interacting systems in nature.

It is very important to mention another crucial point. The interaction scheme in Fig.\ref{xfig3}, \textbf{Theorem}~\ref{theorem-1}, \textbf{Theorem}~\ref{theorem-2} and interaction equations in Eq.\,(\ref{Aydiner}) and all the results we obtained here are valid not only for cosmology but also for all coupled non-equilibrium thermodynamic systems. The results also show that the dynamics of all coupled, interacting, and transforming thermodynamic systems are chaotic both far from and close to the equilibrium.
This is also another significant result of the present study. Additionally, these results may indicate the presence of new thermodynamics laws.

If we sum up, in this study, we theoretically show that the interacting coupling thermodynamic systems behave as chaotic. These theoretical findings will have important contributions to physics. These theoretical discoveries can be proven experimentally, and we expect experimental studies to confirm our theoretical findings.

Finally, we conclude that this result can be generalized to the $N$ coupling systems which behave as interaction networks. In addition to the chaotic behavior of all elements in such an interaction network, the complex system itself is expected to behave chaotically. The human neural network, which is a self-organized system, can be the best example of this. We know that chaos characterizes the dynamics of a system whose dynamics are sensitive to the initial conditions and whose time evolution is unpredictable. However, at this point, we can give a new definition of chaos: \textit{Chaos is the collective minimum action of the synchronized self-organized interacting systems.}

\section{Summary of the Main Results} % R2 version arXiv 19 September 2023
\label{Summary}

\begin{enumerate}
	
	\item  \textbf{Interaction between dark matter and dark energy:} The interaction between dark matter and dark energy is chaotic.
	
	\item  \textbf{Hidden Symmetry:} All coupled interacting particle and thermodynamic systems have hidden symmetries that can be represented by the self-interaction loops.
	
	\item \textbf{Hidden Interaction:} The self-interaction loops of the coupled interacting particle and thermodynamic systems have hidden variables and vectorial fields.
	
	\item \textbf{New Physics  Law:}  The dynamics of all coupled interacting particle and thermodynamic systems are chaotic.
	
	\item \textbf{Definition of the Chaos:} Chaos is the minimum action of the coupled, synchronized, self-organized interacting systems. 
	
	\item \textbf{Self-organization:} All self-organized systems are the manifestation of chaotic behavior.
	
\end{enumerate}

\section{Further Studies} \label{Further}

In this work, we propose a new interaction scheme and new definitions for thermodynamic systems.
Using the scheme, we show that the interactions between dark matter and dark energy must be chaotic. Furthermore, the proposed new interaction scheme offers a new perspective on physics. Therefore, the model may have applications in many areas. On the other hand, one can see that the present study contains many open problems. For instance, we recently examined the interactions between two thermodynamic systems under specific conditions using this scheme \cite{202404.0257}.

In particular, one important open problem is to understand whether the model is compatible with standard gravity theories. Therefore, we are currently resolving this interaction scheme by considering the well-known results of $\Lambda$CDM. One of our most important goals is in Eq.\,(\ref{Aydiner}) to see whether our proposed interaction model will work in accordance with the known results of cosmology. Currently, we are trying to determine the working range of the interaction model by fitting the model to the well-known results of $\Lambda$CDM. Furthermore, we are discussing the scale factor, dark matter and dark energy densities, state parameters and other known cosmological quantities using the model. Similarly, some fundamental problems of cosmology such as singularity, flatness, monopole, transitions between different epochs, entropy, complexity, and the future of the universe are among our immediate goals.

Additionally, discussing the proposed model within the framework of thermodynamics and particle physics and understanding its application areas may provide interesting results. On the other hand, experimental verification of this proposed interaction scheme is of great importance. Studying these open problems may contribute to the development of new insights and is of great importance in terms of the validation of the model.

\section*{Acknowledgements} 

I am grateful to the anonymous referee of IJTP for valuable comments and suggestions that improved the quality of the work. I am grateful to the anonymous referee in the previous submission for the many valuable constructive corrections, suggestions, and contributions that played an important role in improving the manuscript. Finally, I thanks İstanbul University since this work was supported by İstanbul University. Research Project: FYO-2021-38105 which is entitled “Investigation of the Cosmic Evolution of the Universe".

\bibliography{x-Cosmology}

\end{document}